\documentclass[twocolumn,showpacs,preprintnumbers,amsmath,amssymb]{revtex4}

\usepackage{graphicx}
\usepackage{dcolumn}
\usepackage{bm}

\begin{document}


\title{Supersolid helium at high pressure}
\author{E. Kim}
\email{eunseong@kaist.ac.kr}%
\altaffiliation{Department of Physics, Korea Advanced Institute of
Science and Technology, Daejeon, R.O.K}
\author{M. H. W. Chan}%
 \affiliation{%
Department of Physics, the Pennsylvania State University, University Park, Pennsylvania 16802, U.S.A
}%

\date{\today}

\begin{abstract}
We have measured the pressure dependence of the supersolid fraction
by a torsional oscillator technique. Superflow is found from 25.6
bar up to 136.9 bar. The supersolid fraction in the low temperature
limit increases from 0.6 \% at 25.6 bar near the melting boundary up
to a maximum of 1.5\% near 55 bar before showing a monotonic
decrease with pressure extrapolating to zero near 170 bar.
\end{abstract}
\maketitle
Recently we reported the observation of superfluidity in solid
$^4$He confined inside porous Vycor glass with characteristic pore
diameter of 7 nm \cite{vycorto}, porous gold with pore diameter of
490 nm \cite{pgto} and also in bulk solid $^4$He \cite{bulkto}. In
the bulk experiment solid helium is confined in an annular channel
inside the torsion bob (see inset \textbf{III } of Fig.1). The
width, depth and outer diameter of the channel are respectively 0.63
mm, 5 mm, and 10 mm. The torsion bob is driven and maintained at a
resonant period of 1,096,465 ns with a mechanical quality factor
($Q$) of $2 \times 10^6$ by a pair of electrodes. When the torsional
oscillator is cooled below 230 mK the resonant period (which is
proportional to the squared-root of the moment of inertia, $I$, of
the torsion bob) shows an abrupt drop from the expected, linearly
extrapolated value from higher temperatures. The most simple
explanation of the decrease in $I$, considering the various control
experiments that were carried out, is the onset of nonclassical
rotational inertia (NCRI) or superfluidity in solid $^4$He \cite{ss4}.%
\begin{figure}[b]
\includegraphics[%
width=0.7\columnwidth]{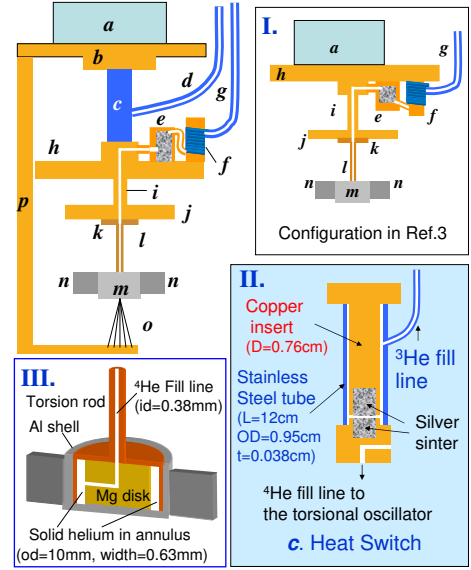} \caption{\label{fig:epsart}
Experimental configurations. The insets depict \textbf{I.}
configuration in Ref.3,\textbf{ II.} details of heat switch, and
\textbf{III.} annular channel in the cell. \textbf{\emph{a}}. mixing
chamber, \textbf{\emph{b}}. thermal platform, \textbf{\emph{c}}.
$^3$He heat switch, \textbf{\emph{d}}. $^3$He fill line,
\textbf{\emph{e}}. silver sinter heat sink, \textbf{\emph{f}}. wound
capillary heat sink, \textbf{\emph{g}}. $^4$He fill line,
\textbf{\emph{h}}. thermal platform, \textbf{\emph{i}}. vibration
isolator, \textbf{\emph{j}}. thermal platform, \textbf{\emph{k}}.
torsional oscillator base, \textbf{\emph{l}}. torsion rod,
\textbf{\emph{m}}. torsion cell, \textbf{\emph{n}}. two electrodes,
\textbf{\emph{o}}. copper wires, \textbf{\emph{p}}. cold finger }
\end{figure}%

The supersolid fraction in the low temperature limit found in the
Vycor and the bulk experiments is on the order of the 1\% in spite
of vastly different surface to volume ratio (a factor of
$2.5\times10^{4}$) of the space available for helium. This indicates
the observed superfluidity is not a surface related phenomenon. It
is also difficult to reconcile this observation with the suggestion
that the superfluidity is due entirely to defects, dislocations, and
other imperfections in the crystal since this would require the
crystallite size in the bulk sample to be the same as that in Vycor
or at most 7 nm.

Nevertheless, a number of theoretical papers suggest that
superfluidity is unlikely to occur in a perfect crystal
\cite{ss1,ceperley1,nikola,buro}. Andreev and Liftshitz suggested in
1969 a specific scenario that Bose condensation of zero point
vacancies and other defects can lead to superfluidity in solid
helium \cite{ss2}. If the observed superflow is in fact a simple
consequence of condensation of zero point vacancies then the
supersolid fraction should decrease as the pressure (and hence
density) of the solid sample is increased away from the melting
boundary deep into the solid phase. In the bulk solid experiment a
total of 17 samples of solid helium with pressure ranging from 26
bar to 65 bar were studied \cite{bulkto}. While superflow was found
in every sample, the value of the supersolid fraction in the low
temperature limit was found to vary between 0.6 and 1.7\% with no
obvious dependence on pressure (Fig. 4).

The set of measurements reported below was undertaken to understand
and to reduce the scatter in the supersolid fraction. We made the
assumption that the scatter in the supersolid fraction is a
consequence of the random orientation of small crystallites inside
the annular channel of the torsion cell. Solid helium has been grown
from liquid under constant pressure \cite{cp1, cp3, cp2, cp4, cp5},
constant temperature \cite{ct1, ct2, ct3, ct4}, and constant volume
methods \cite{xray, tc, neutron1, neutron2, neutron3}. It has been
reported that the constant pressure method tends to grow a solid
sample with high degree of crystallinity \cite{cp1, cp2}. Growing
solid from the superfluid phase at a constant temperature also
results in a high quality crystal \cite{ct3, ct4}, but the pressure
of the samples is limited to below 30 bar. There are reports that
the constant volume method also results in solid samples of
reasonable quality \cite{xray, tc, neutron1, neutron2, neutron3}.
There is consensus, however, the crystal quality of solid grown
under constant volume condition is inferior to the other two methods
\cite{cp1,ct3}. The solid samples studied in Ref. \cite{bulkto} were
grown using the blocked capillary (constant volume) method. In this
method helium under high pressure is introduced from room
temperature into a sample cell via a thin capillary. The capillary
is thermally anchored at different points inside the cryostat of
progressively lower temperatures. After liquid helium of the desired
density has been introduced into the sample cell, the temperature of
the capillary at a certain point is lowered to solidify the helium
within and form a plug. The liquid in the constant volume below the
solid plug, including the sample cell can then be cooled into the
solid phase with a concomitant drop in pressure.

The experimental configuration of Ref. \cite{bulkto} is shown in
inset \textbf{I}, Fig. 1. Liquid helium from the capillary
(\emph{\textbf{g}}) is introduced from the base (\textbf{\emph{k}})
of the oscillator through the torsion rod (\textbf{\emph{l}}) to the
torsion cell (\textbf{\emph{m}}). The base of the oscillator was
attached to a thermal platform (\textbf{\emph{j}}) which was in turn
connected to the mixing chamber (\textbf{\emph{a}}) of the dilution
refrigerator through another platform (\textbf{\emph{h}}). Platforms
\textbf{\emph{h}} and \textbf{\emph{j}} are connected through a
vibration isolator (\textbf{\emph{i}}) in the form of a hollow
copper cylinder. In such a configuration, liquid helium in the
torsion rod will freeze first. The solidification will then proceed
through the long narrow hole (i.d.=0.38 mm and length=7.5 mm)
drilled inside the magnesium disk (Fig.1, inset \textbf{III}) before
reaching the annulus of channel width=0.63 mm. It is therefore not
surprising that such a solidification process through this long
narrow path will result in polycrystalline samples with grains of
size no larger than 0.38 mm with random and un-reproducible
orientations inside the annulus. We think this unfavorable growth
process of the solid is the primary reason for the scatter in the
supersolid fraction.
\begin{figure}[b]
\includegraphics[%
width=0.85\columnwidth ]{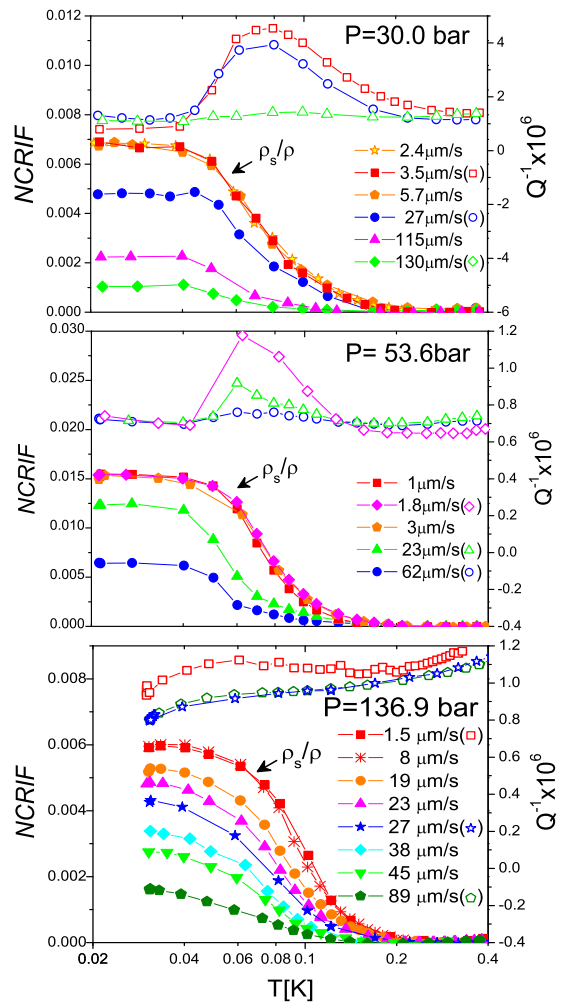} \caption{\label{fig:epsart}$NCRIF$
(closed symbols) and dissipation (in $Q^{-1}$, open symbols) as a
function of temperature for three samples at P=30, P=53.6, and
P=136.9 bar. The increases of the resonant period due to the filling
of the cell are 2814, 3045, and 3700 ns respectively. $NCRIF$
measured at the limit of low oscillation speed corresponds to
$\rho_{s}/\rho$. Broad dissipation peak is seen for data taken at
low oscillation speed in low pressure samples but barely discernible
at 136.9 bar}
\end{figure}

In the current experiment, we used the same torsional oscillator as
that in Ref. \cite{bulkto}. However, we have installed a heat switch
(Fig.1 inset \textbf{II}) between thermal platform \emph{\textbf{h}}
and the mixing chamber in an attempt to change the cooling path of
the torsion cell during the growth of solid to facilitate the
nucleation of solid helium from the bottom of the annulus. The heat
switch can be opened (closed) by emptying (filling) the thin wall
stainless steel tubing with liquid $^3$He. When a solid is being
grown in the torsion cell, the heat switch is opened and the latent
heat of freezing is designed to be primarily carried from the
torsion cell to the mixing chamber through 10 strands of 0.05 mm
diameter copper wires (\emph{\textbf{o}}) attached to the bottom of
the torsion cell. The other ends of the wires are attached to a
heavy copper bar (\emph{\textbf{p}}) that is firmly anchored to the
mixing chamber (Fig.1).

During the growth of solid helium the pressure and the density of
the sample in the cell were monitored by a resistance strain gauge
(glued onto the wall of the torsion cell) and the increase in the
period of the oscillator. In all solid samples we have grown for
this study, we found the resonant period always shows an increase
before the pressure showed any noticeable decrease. This indicates
that initially solid nucleates under the constant pressure growth
condition and it is reasonable to speculate that the nucleation
starts at the bottom of the annulus, close to the copper wires. In
spite of repeated effort, we found it impossible to complete the
solidification process under the constant pressure condition. What
we found is that before the growth of solid in the torsion cell is
completed the pressure always exhibits a drop indicating a solid
block is formed cutting off the supply of helium into the torsion
cell. It likely occurs when the solid in the annulus grows into the
narrow hole in the magnesium disk. The copper wires were sufficient
to cool the torsional oscillator down to 1.3 K. Further cooling
requires the introduction of liquid $^3$He into the heat switch. The
pressure readings of the solid helium samples are consistent with
the density deduced by measuring the change in the resonant period
from the empty cell value.

\begin{figure}[b]
\includegraphics[%
width=0.85\columnwidth]{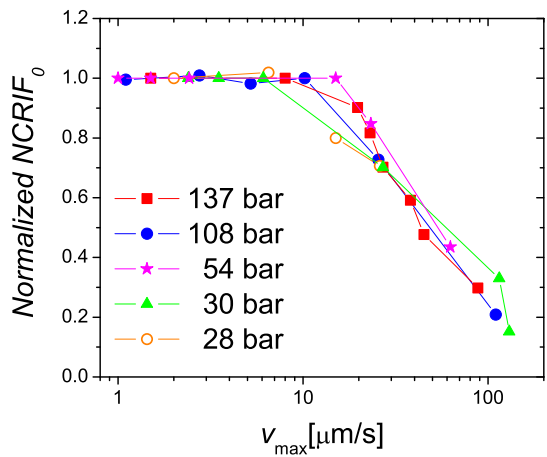} \caption{\label{fig:epsart}
Oscillation speed dependence of normalized $NCRIF$ for five solid
helium samples of $P$=28.1, 30, 53.6, 108.3, and 136.9 bar.
$v_{max}$ is the maximum speed of the annulus holding solid $^4$He.
$NCRIF$ is normalized by the low temperature supersolid fraction,
$\rho_{so}/ \rho$. $\rho_{so}/\rho$ for each sample are 0.00691,
0.00679, 0.01544, 0.00612, and 0.00576 respectively. }
\end{figure}
\begin{figure}[b]
\includegraphics[width=0.85\columnwidth]{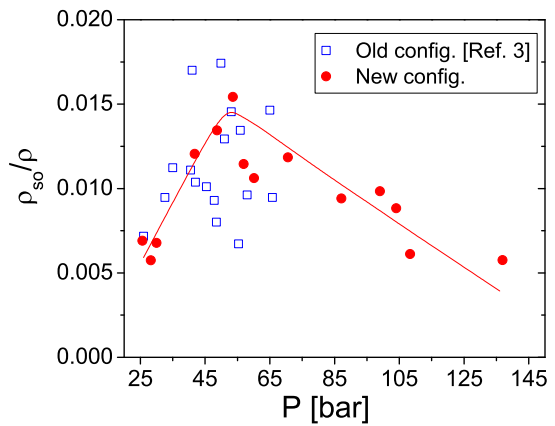}
\caption{\label{fig:epsart}Supersolid fraction in the low
temperature limit, $\rho_{so}/\rho$, as a function of pressure.
Solid samples prepared with heat switch (new configuration) yield
data with less scatter. The solid line is a guide to the eyes.}
\end{figure}
We have grown and studied 14 solid samples with pressure ranging
from 25.6 to 136.9 bar. All samples, including the sample at 136.9
bar, exhibit supersolid decoupling. The temperature dependence of
the supersolid fraction of each sample resembles those in Ref.
\cite{bulkto}. Fig. 2 shows the nonclassical rotational inertia
fraction ($NCRIF$) as a function of temperature at different maximum
oscillation speed of the annulus, $v_{max}$, for solid samples of
30.0, 53.6, and 136.9 bar. The $NCRIF$ results are deduced from the
resonant period following the same procedures outlined in Ref.
\cite{bulkto}. The reproducibility of the resonant period readings
is about 0.5 ns. The resonant period increases on the order of 3000
ns with the filling of the solid sample. This translates an error
bar in $NCRIF$ and $\rho_s/\rho$ to about $2 \times 10^{-4}$. Since
the determination of $NCRIF$ involves the subtraction of the
measured period from a temperature dependent background curve, there
may be an additional systematic error in $NCRIF$ (and $\rho_s/\rho$)
of comparable magnitude. The dissipation, in inverse quality factor
($Q^{-1}$) of the oscillator, deduced from the amplitude of
oscillation at three different $v_{max}$ of each sample are also
shown with open symbols. Broad maxima centering near where $NCRIF$
is changing most rapidly are found. These broad maxima in
dissipation are more pronounced in low pressure solid samples and in
data taken at low $v_{max}$. The dissipation maximum fades with
increasing pressure and it is barely discernible in samples with pressure exceeding 108 bar. %

Fig. 3 shows normalized $NCRIF_0$ (the low temperature limit of
$NCRIF$ at different $v_{max}$ divided by its value obtained at the
lowest $v_{max}$) as a function of $v_{max}$ for solid samples at
five different pressures. These five sets of data show a much better
'collapse' onto a single curve compared to the data shown in Fig. 3
(panel D) of Ref. \cite{bulkto}. The $NCRIF$  is independent of
$v_{max}$, provided $v_{max}$ does not exceed 10 $\mu m/s$. Once
exceeded, $NCRIF$ decreases with $v_{max}$. We interpret this as a
critical velocity effect and as noted in Ref. \cite{bulkto}, the
result indicates that superflow in solid helium becomes dissipative
with the appearance of a single vortex with unity quantum
circulation (if the effective mass is a third of the atomic mass) or
just a few vortices.

$NCRIF$ measured with $v_{max}$ smaller than 10 $\mu m/s$, being
independent of oscillation speed, represents the supersolid
fraction, $\rho_{s}/\rho$. We have used oscillation speed of $5 \mu
m/s$ or less to study the supersolid response of 9 additional solid
samples at 25.6, 41.8, 48.7, 56.9, 60.1, 70.6, 87.1, 99.0, and 104.0
bar. The uncertainty in pressure determination is less than 0.5 bar.
The low temperature supersolid fractions, $\rho_{so}/\rho$, of all
fourteen samples are plotted in Fig. 4 as a function of pressure.
The new data show the supersolid fraction increases from 0.6\% near
the melting pressure up to a maximum of 1.5\% near 55 bar before
decreasing with further increase in pressure. A linear extrapolation
suggests the supersolid fraction will be reduced to zero for
pressures exceeding 170 bar. Unfortunately the torsion cell exploded
as we attempted to make a solid sample of 170 bar.

While the data taken with the new experimental configuration appears
to be much improved over those shown in Ref. [3], the point to point
scatter of the $\rho_{so}/\rho$ values as shown in Fig. 4 is
typically 0.15\% and for the three data points near 55 bar it is as
large as 0.5\%.  These values are many times larger than the
uncertainty in $\rho_{s}/\rho$ obtained in an individual sample.
This suggested that we have not been growing solid samples in a
completely reproducible manner and there is still substantial
variation in the 'crystal quality' of these solid samples.
Measurements on solid samples contained in a torsion cell with
simple cylindrical geometry without an annulus and grown entirely
under the constant pressure may reduce the scatter further.

The non-monotonic dependence of the supersolid fraction on pressure
indicates that, as noted above, the origin of supersolidity is more
subtle than just the simple Bose condensation of zero point
vacancies. The fact that we found a supersolid fraction of up to
1.5\% is also difficult to reconcile with the simple vacancy
condensation model. A number of experiments \cite{bernier,simmons}
give indirect evidence that zero point vacancies, if present below
0.2K, would be much smaller than 1\% of the lattice sites.

Solid helium at an elevated pressure is expected to be less quantum
mechanical than that at a lower pressure \cite{nosanow,nosanow2}.
X-ray diffraction studies measuring the zero-point energy induced
motions of the $^4$He atoms from their lattice sites appear to
confirm this expectation \cite{cp4,ro2}. The declining supersolid
fraction with pressure beyond 55 bar is also consistent with this
expectation. However, we do not understand why there is no apparent
change in $T_c$ with pressure.

It has been suggested that a perfect solid helium crystal cannot
support superflow \cite{ceperley1,nikola,buro}. This idea received
support from the recent torsional oscillator experiment of Rittner
and Reppy \cite{reppy}. They found supersolid decoupling in a solid
sample made by the same blocked capillary method. However, upon
annealing the sample by cooling it much more slowly from about 1.5K
than when it was first grown, the supersolid decoupling is found to
diminish and even disappear. We have looked for this annealing
effect by cooling a number of solid samples from the liquid-solid
coexistence region down to the lowest temperature at a cooling rate
that is up to 5 times slower than that of Rittner and Reppy. We
found the supersolid fraction due to different annealing procedure
can differ by at most 15\%. We have not been able to eliminate the
superflow in any of the more than 50 bulk solid samples we have
studied so far in our laboratory. In addition to Rittner and Reppy,
our observation of superflow in solid helium with the torsional
oscillator technique was also replicated by the Shirahama group at
Keio University \cite{shirahama} and the Kubota group of the
University of Tokyo \cite{kubota}. These two groups have also tried
but failed to eliminate superflow by annealing.%

To conclude, we observed the supersolid phase extends at least up to
136.9 bar. The supersolid fraction appears to increase with pressure
from the melting pressure up to 55 bar and then decreases with
further increase with pressure. Linear extrapolation indicates the
supersolid phase terminates near 170 bar.  The critical velocity of
the superflow is found to be on the order of $10 \mu m/s$.
We want to acknowledge informative discussions with J. Beamish, A.
C. Clark, H. Kojima, M. Kubota, X. Lin, J. D. Reppy, K. Shirahama,
and J. T. West. The work is supported by NSF under grant DMR
0207071. One of us, M. H. W. C. would like to thank the Kavli
Institute for Theoretical Physics for hospitality during the
supersolid workshop.

\end{document}